\documentclass[aps,prd,twocolumn,superscriptaddress,letterpaper,floatfix,nofootinbib,preprintnumbers]{revtex4}
\usepackage{epsfig,units,color,ulem}

\newcommand{\at}{\alpha_{T}}
\renewcommand{\>}{\ensuremath{\!>}}

\begin{document}
\preprint{FERMILAB-PUB-15-102-ND}

\title{Observation of seasonal variation of atmospheric
multiple-muon events in the MINOS Near and Far Detectors}
\newcommand{\Berkeley}{Lawrence Berkeley National Laboratory, Berkeley, California, 94720 USA}
\newcommand{\Cambridge}{Cavendish Laboratory, University of Cambridge, Madingley Road, Cambridge CB3 0HE, United Kingdom}
\newcommand{\Cincinnati}{Department of Physics, University of Cincinnati, Cincinnati, Ohio 45221, USA}
\newcommand{\FNAL}{Fermi National Accelerator Laboratory, Batavia, Illinois 60510, USA}
\newcommand{\RAL}{Rutherford Appleton Laboratory, Science and TechnologiesFacilities Council, Didcot, OX11 0QX, United Kingdom}
\newcommand{\UCL}{Department of Physics and Astronomy, University College London, Gower Street, London WC1E 6BT, United Kingdom}
\newcommand{\Caltech}{Lauritsen Laboratory, California Institute of Technology, Pasadena, California 91125, USA}
\newcommand{\Alabama}{Department of Physics and Astronomy, University of Alabama, Tuscaloosa, Alabama 35487, USA}
\newcommand{\ANL}{Argonne National Laboratory, Argonne, Illinois 60439, USA}
\newcommand{\Athens}{Department of Physics, University of Athens, GR-15771 Athens, Greece}
\newcommand{\NTUAthens}{Department of Physics, National Tech. University of Athens, GR-15780 Athens, Greece}
\newcommand{\Benedictine}{Physics Department, Benedictine University, Lisle, Illinois 60532, USA}
\newcommand{\BNL}{Brookhaven National Laboratory, Upton, New York 11973, USA}
\newcommand{\CdF}{APC -- Universit\'{e} Paris 7 Denis Diderot, 10, rue Alice Domon et L\'{e}onie Duquet, F-75205 Paris Cedex 13, France}
\newcommand{\Cleveland}{Cleveland Clinic, Cleveland, Ohio 44195, USA}
\newcommand{\Delhi}{Department of Physics \& Astrophysics, University of Delhi, Delhi 110007, India}
\newcommand{\GEHealth}{GE Healthcare, Florence South Carolina 29501, USA}
\newcommand{\Harvard}{Department of Physics, Harvard University, Cambridge, Massachusetts 02138, USA}
\newcommand{\HolyCross}{Holy Cross College, Notre Dame, Indiana 46556, USA}
\newcommand{\Houston}{Department of Physics, University of Houston, Houston, Texas 77204, USA}
\newcommand{\IIT}{Department of Physics, Illinois Institute of Technology, Chicago, Illinois 60616, USA}
\newcommand{\Iowa}{Department of Physics and Astronomy, Iowa State University, Ames, Iowa 50011 USA}
\newcommand{\Indiana}{Indiana University, Bloomington, Indiana 47405, USA}
\newcommand{\ITEP}{High Energy Experimental Physics Department, ITEP, B. Cheremushkinskaya, 25, 117218 Moscow, Russia}
\newcommand{\JMU}{Physics Department, James Madison University, Harrisonburg, Virginia 22807, USA}
\newcommand{\LASL}{Nuclear Nonproliferation Division, Threat Reduction Directorate, Los Alamos National Laboratory, Los Alamos, New Mexico 87545, USA}
\newcommand{\Lebedev}{Nuclear Physics Department, Lebedev Physical Institute, Leninsky Prospect 53, 119991 Moscow, Russia}
\newcommand{\LLL}{Lawrence Livermore National Laboratory, Livermore, California 94550, USA}
\newcommand{\LosAlamos}{Los Alamos National Laboratory, Los Alamos, New Mexico 87545, USA}
\newcommand{\Manchester}{School of Physics and Astronomy, University of Manchester, Oxford Road, Manchester M13 9PL, United Kingdom}
\newcommand{\MIT}{Lincoln Laboratory, Massachusetts Institute of Technology, Lexington, Massachusetts 02420, USA}
\newcommand{\Minnesota}{University of Minnesota, Minneapolis, Minnesota 55455, USA}
\newcommand{\Crookston}{Math, Science and Technology Department, University of Minnesota -- Crookston, Crookston, Minnesota 56716, USA}
\newcommand{\Duluth}{Department of Physics, University of Minnesota Duluth, Duluth, Minnesota 55812, USA}
\newcommand{\Ohio}{Center for Cosmology and Astro Particle Physics, Ohio State University, Columbus, Ohio 43210 USA}
\newcommand{\Otterbein}{Otterbein College, Westerville, Ohio 43081, USA}
\newcommand{\Oxford}{Subdepartment of Particle Physics, University of Oxford, Oxford OX1 3RH, United Kingdom}
\newcommand{\PennState}{Department of Physics, Pennsylvania State University, State College, Pennsylvania 16802, USA}
\newcommand{\PennU}{Department of Physics and Astronomy, University of Pennsylvania, Philadelphia, Pennsylvania 19104, USA}
\newcommand{\Pittsburgh}{Department of Physics and Astronomy, University of Pittsburgh, Pittsburgh, Pennsylvania 15260, USA}
\newcommand{\IHEP}{Institute for High Energy Physics, Protvino, Moscow Region RU-140284, Russia}
\newcommand{\Rochester}{Department of Physics and Astronomy, University of Rochester, New York 14627 USA}
\newcommand{\RoyalH}{Physics Department, Royal Holloway, University of London, Egham, Surrey, TW20 0EX, United Kingdom}
\newcommand{\Carolina}{Department of Physics and Astronomy, University of South Carolina, Columbia, South Carolina 29208, USA}
\newcommand{\SDakota}{South Dakota School of Mines and Technology, Rapid City, South Dakota 57701, USA}
\newcommand{\SLAC}{Stanford Linear Accelerator Center, Stanford, California 94309, USA}
\newcommand{\Stanford}{Department of Physics, Stanford University, Stanford, California 94305, USA}
\newcommand{\StJohnFisher}{Physics Department, St. John Fisher College, Rochester, New York 14618 USA}
\newcommand{\Sussex}{Department of Physics and Astronomy, University of Sussex, Falmer, Brighton BN1 9QH, United Kingdom}
\newcommand{\TexasAM}{Physics Department, Texas A\&M University, College Station, Texas 77843, USA}
\newcommand{\Texas}{Department of Physics, University of Texas at Austin, 1 University Station C1600, Austin, Texas 78712, USA}
\newcommand{\TechX}{Tech-X Corporation, Boulder, Colorado 80303, USA}
\newcommand{\Tufts}{Physics Department, Tufts University, Medford, Massachusetts 02155, USA}
\newcommand{\UNICAMP}{Universidade Estadual de Campinas, IFGW-UNICAMP, CP 6165, 13083-970, Campinas, SP, Brazil}
\newcommand{\UFG}{Instituto de F\'{i}sica, Universidade Federal de Goi\'{a}s, CP 131, 74001-970, Goi\^{a}nia, GO, Brazil}
\newcommand{\USP}{Instituto de F\'{i}sica, Universidade de S\~{a}o Paulo,  CP 66318, 05315-970, S\~{a}o Paulo, SP, Brazil}
\newcommand{\Warsaw}{Department of Physics, University of Warsaw, Pasteura 5, PL-02-093 Warsaw, Poland}
\newcommand{\Washington}{Physics Department, Western Washington University, Bellingham, Washington 98225, USA}
\newcommand{\WandM}{Department of Physics, College of William \& Mary, Williamsburg, Virginia 23187, USA}
\newcommand{\Wisconsin}{Physics Department, University of Wisconsin, Madison, Wisconsin 53706, USA}
\newcommand{\deceased}{Deceased.}

\affiliation{\ANL}
\affiliation{\BNL}
\affiliation{\Caltech}
\affiliation{\Cambridge}
\affiliation{\UNICAMP}
\affiliation{\Cincinnati}
\affiliation{\FNAL}
\affiliation{\UFG}
\affiliation{\Harvard}
\affiliation{\HolyCross}
\affiliation{\Houston}
\affiliation{\IIT}
\affiliation{\Indiana}
\affiliation{\Iowa}
\affiliation{\UCL}
\affiliation{\Manchester}
\affiliation{\Minnesota}
\affiliation{\Duluth}
\affiliation{\Otterbein}
\affiliation{\Oxford}
\affiliation{\Pittsburgh}
\affiliation{\RAL}
\affiliation{\USP}
\affiliation{\Carolina}
\affiliation{\Stanford}
\affiliation{\Sussex}
\affiliation{\TexasAM}
\affiliation{\Texas}
\affiliation{\Tufts}
\affiliation{\Warsaw}
\affiliation{\WandM}

\author{P.~Adamson}
\affiliation{\FNAL}


\author{I.~Anghel}
\affiliation{\Iowa}
\affiliation{\ANL}



\author{A.~Aurisano}
\affiliation{\Cincinnati}









\author{G.~Barr}
\affiliation{\Oxford}









\author{M.~Bishai}
\affiliation{\BNL}

\author{A.~Blake}
\affiliation{\Cambridge}


\author{G.~J.~Bock}
\affiliation{\FNAL}


\author{D.~Bogert}
\affiliation{\FNAL}




\author{S.~V.~Cao}
\affiliation{\Texas}

\author{C.~M.~Castromonte}
\affiliation{\UFG}




% ...

\collaboration{The MINOS Collaboration}
\noaffiliation

\noaffiliation

\date{\today}

\begin{abstract}


We report the first observation of seasonal modulations in 
the rates of cosmic ray multiple-muon
events at two underground sites, the MINOS Near 
Detector with an overburden of 225 mwe, and the
MINOS Far Detector site at
2100 mwe. At the deeper site, multiple-muon events
with muons separated by more than 8 m exhibit 
a seasonal rate that peaks during the summer,  
similar to that of single-muon events.
In contrast and unexpectedly,
the rate of multiple-muon events with muons 
separated by less than 5-8 m, and the rate of multiple-muon
events in the smaller, shallower Near Detector, 
exhibit a seasonal rate modulation that peaks in the winter.
\\

\end{abstract}

\keywords{MINOS \sep Atmospheric Muons \sep Seasonal Variations \sep Multimuons  \sep charge ratio}

\maketitle

\section{Introduction}
\label{Introduction}
Muons observed in underground particle 
detectors originate from the 
interactions of cosmic rays with nuclei in the upper atmosphere.  These 
interactions produce pions ($\pi$) and kaons ($K$) which can either
interact, generating hadronic cascades, or decay, producing muons.  
The probability that these mesons will decay rather than interact 
is dependent on their energy and the density of the atmosphere near 
their point of production. 
The temperature of the upper atmosphere varies slowly over the 
year, causing a seasonal effect on underground muon rates.
 Increases in the temperature of the atmosphere decrease the
local density and thus reduce the 
probability that a secondary meson will interact. Consequently, the 
muon flux should increase in the summer.  
A number of experiments have observed this variation in the single muon
rate~\cite{Barret:1952,Castagnoli:1967,Sherman:1954,Torino:1967,Hobart:1961,
Baksan:1987,Ambrosio:1997tc,Bouchta:1999,Bellini:2012te,Solvi:2009,
Desiati:2011}, including MINOS in both 
Far Detector (FD) data~\cite{Osprey:2009ni,Adamson:2009zf} 
and Near Detector (ND) data~\cite{Adamson:2014}. 

Seasonal variations for single muons have  been studied with a
correlation coefficient $\at$ defined by:
\begin{equation}
\frac{\Delta R_{\mu}}
{\<R_{\mu}\>}=\at\frac{\Delta T_{eff}}{\<T_{eff}\>}
\label{eq:alpha}
\end{equation}
where $\<R_{\mu}\>$ is the mean muon rate, and is equivalent to the 
rate for an effective atmospheric temperature equal to $\<T_{eff}\>$. The 
magnitude of the temperature coefficient $\at$ is dependent 
on the muon energy at production and hence
the depth of the detector. 
The effective temperature $T_{eff}$ is a weighted average over
the region of the atmosphere where the muons originate.

By the same reasoning as above a 
variation should also be present in the rate of multiple-muon events. 
No such studies of multiple-muon seasonal 
rates are reported in the literature.
The formulae used to calculate $T_{eff}$ for single muons
assume a single leading hadron from the first interaction is the parent,
an assumption that is not applicable for multiple-muon events.

The probability that a cosmic ray shower will give a multiple-muon
event observed in the MINOS Near or Far detectors 
is enhanced whenever any of the following conditions are true:
1)  The primary interaction occurs high in the
atmosphere where the density is lower and a larger
fraction of produced hadrons decay;
2)  The energy of the primary is large so a higher multiplicity
of hadrons is produced; 
3) The
cosmic ray  primary is a heavy nucleus which breaks up and makes
more hadrons;
and 4) A leading hadron decays to dimuons.
Assuming the relative probability of interaction and decay for each
meson in a shower is independent, for multiple muons that come 
from the same energy and altitude as
a single muon event, one might expect
an increase in rate during the summer that is roughly proportional
to the muon multiplicity, N, such that
 $\alpha_{T,N}$ = N $\times$ $\alpha_{T,1}$.  The result
presented here differs greatly from this.
This paper presents the first measurement of the 
multiple-muon modulation parameters.

Note that most extensive air showers have many muons in them, 
but that the highest
energy muons which can reach an underground detector are
produced in the first few interactions.  Observed single-muon 
events are most likely multiple muons in 
which any other muons range out before 
reaching the detector or missed the detector laterally.  
A single muon observed in a detector underground is most likely
the highest energy muon from the shower 
due to the steeply falling cosmic ray energy spectrum.

The 
MINOS detectors and the event selection are 
described in Sec.~\ref{sec:MINOS}.
In Sec.~\ref{sec:Modulation} the measurement and comparison of the 
modulation parameters for the MINOS ND and FD multiple-muon 
and single-muon event rates are presented. In Sec.~\ref{sec:Discussion} and 
Sec.~\ref{sec:Conclusion} some possible explanations of the 
seasonal behavior of the multiple-muon rates are considered.

\section{The MINOS Detectors and Muon Data}
\label{sec:MINOS}

The MINOS detectors are planar magnetized steel/scintillator tracking
calorimeters \cite{Michael:2008bc}. The vertically oriented detector
planes are composed of \unit[2.54]{cm} thick steel and \unit[1]{cm} thick
plastic scintillator. A scintillator layer is composed of
\unit[4.1]{cm} wide strips. The MINOS ND has a total mass of \unit[0.98]{kton},
and lies \unit[104]{m} (225 mwe) 
underground at Fermilab at 42$^\circ$ North latitude. The detector
is made from \unit[3.8]{m} $\times$ \unit[4.8]{m} hexagonal planes and is 
\unit[17]{m} long.  It consists of two sections, a calorimeter encompassing the
upstream 121 planes and a spectrometer containing the
downstream 161 planes. In both sections, one out of every
five planes is covered with 96 scintillator strips
attached to the steel planes. In the calorimeter section, the
other four out of five planes are covered with 64
scintillator strips, while in the spectrometer section they
have no scintillator.
Only muons which enter the calorimeter are included in this 
analysis.  The larger FD is
\unit[705]{m} below the surface (2100 mwe), has a total mass
of \unit[5.4]{kton}, and is located in the Soudan
Underground Laboratory, at 48$^\circ$ North latitude. It is
composed of 484 steel-scintillator \unit[8.0]{m} octagonal planes 
and is \unit[31]{m} long.  The detectors are oriented to face 
the NuMI beam, but through-going cosmic muons are well reconstructed 
over wide geometric angular regions.

Six {years} of MINOS ND data collected 
between June~1,~2006 and April 30,~2012
and \unit[9]{years} of MINOS FD data collected between 
August~1,~2003 and April~30,~2012 are analyzed for this paper.
The cosmic muon trigger criteria are similar at both detectors requiring 
that a signal is registered in
either 4 strips in 5 sequential planes 
or that strips from any  20 planes register a total signal 
above threshold within a given time window. The raw cosmic trigger rate 
at the ND and FD are approximately \unit[27]{Hz} 
and \unit[0.5]{Hz} respectively.

\begin{figure}[thb]
  \begin{center}
    \includegraphics[width=0.48\textwidth]{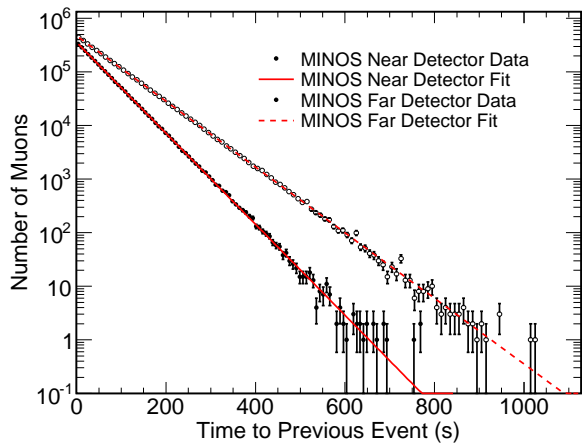}
  \end{center}
  \caption{Time between neighboring atmospheric multiple-muon events 
in the MINOS detectors. The data are well described by an exponential
over six orders of magnitude in instantaneous rate.}
  \label{fig:DeltaTmulti}
\end{figure}

The single-muon event selection requires there to be a 
single reconstructed track in an event.
The multiple-muon event selection requires there to be 
more than one reconstructed track in an event.
However, since the single-muon event rate is much larger than the 
multiple-muon event rate, the multiple-muon sample contains
a background of single-muon events that have been
mis-reconstructed to contain two tracks
This background 
is greatly reduced by requiring that, for multi-track events, the 
track separation, $\Delta S$, defined as the minimum point of closest
approach between any two tracks, 
be greater than \unit[0.6]{m}.
Observed excesses due to this background at small $\Delta S$ in
both the ND and FD were removed by this selection, 
reducing the background 
from 1.3\% to less than 200 events out of 11 million in the FD.
Figure~\ref{fig:DeltaTmulti} 
shows the time between sequential multiple-muon events. The 
multiple-muon event rates at the ND and FD are  
\unit[19.6]{mHz} and \unit[14.1]{mHz} respectively. In total the 
MINOS ND and FD  have 
collected 2.45$\times$10$^{6}$ and 3.36$\times$10$^{6}$ good 
multiple-muon events respectively.

\begin{figure}[thb]
  \begin{center}
    \includegraphics[width=0.48\textwidth]{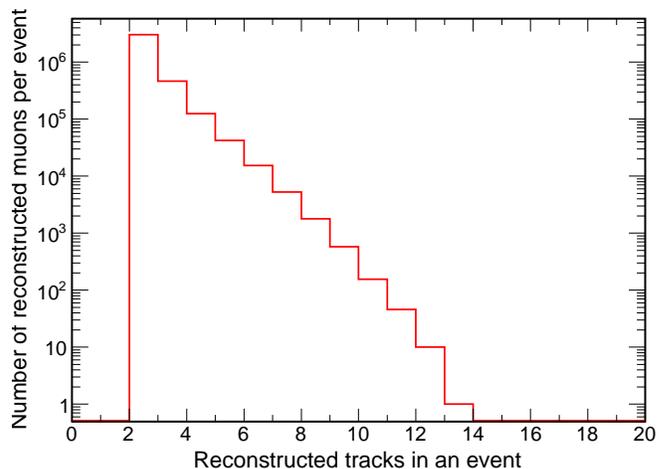}
  \end{center}
  \caption{The reconstructed muon multiplicity,
for events containing more than one reconstructed track,
in the Far Detector.}
  \label{fig:multiplicity}
\end{figure}

The rate of multiple muons in the MINOS detectors is dominated by
   m$_\mu$ = 2 and m$_\mu$ = 3, where m$_\mu$ is the muon multiplicity.
   For the FD, the reconstruction works well for these small multiplicities,
   identifying all tracks in 84\% (67\%) of events for m$_\mu$ = 2 (3).
   From a multiple-muon Monte Carlo \cite{bib:cesar},
the efficiency for identifying an event as a multiple muon is
   84\% (94\%) for m$_\mu$ = 2 (3), rising to above 97\% for 
m$_\mu$~$>$~3.
   However, the reconstructed multiplicity is frequently too low
   for high multiplicity events. No event with m$_\mu$~$>$~13 is
   recorded in either of the MINOS detectors. 
Similar measurements with the finer-grained Soudan 2 detector at the
same depth as the MINOS FD recorded 
multiplicities up to 20 \cite{bib:kasahara}.  In the coarser-grained MINOS
detectors, the individual tracks closest in distance from such 
high-multiplicity events will be resolved as a single track or not 
pass the track quality criteria used in the MINOS reconstruction
algorithms.  
   Figure~\ref{fig:multiplicity}
 shows the reconstructed multiplicity distribution
   in the MINOS FD.

\section{Modulation Analysis} 
\label{sec:Modulation}
To compare the variation in the event rates for multiple-muon 
and single-muon events, the rates are fit to a 
sinusoidally-varying function of time.  There is
no a-priori 
reason to believe that the rates vary sinusoidally
through the year, but this fit gives a 
qualitatively useful amplitude and phase.  
The following function, which contains four free parameters,
is used for the fit:
\begin{equation}
R(t)=R_{0}(1-\frac{f t}{365.25})(1+A 
\cos[\frac{2\pi}{T}(t-t_{0})])
\label{eq:fit}
\end{equation}
where t is the number of days since Jan.~1,~2010 and $t_{0}$ 
is the phase; $R_{0}$ is the mean rate on Jan.~1,~2010; $A$ is the 
modulation amplitude and T is the period (approximately 1 year).  
The parameter $f$ is the loss rate 
(described in
Reference \cite{Adamson:2014}) that accounts for an observed linear
decrease
 in the event
rate in both the FD and ND over the lifetime of the experiment.  
The source of this small
but apparently steady decrease has not been conclusively identified
and is under study.
The best-fit parameters 
are given in Table~\ref{tab:MultiModulation}.

\begin{table*}[htb]
\begin{tabular}{lcccc}\hline
Data Set   & Amplitude & Loss Rate (f) & Period (T) & Phase (t$_0$)\\ 
           & (\%)      & (\%/year) & (days) & (days) \\ \hline \hline
            \multicolumn{5}{c}{MINOS FD } \\ \hline
$\Delta$S$>$\unit[0.6]{m}               & 0.39$\pm$0.08 & -0.04$\pm$0.02  & 356.4$\pm$4.1 & 105.2$\pm$16.1\\ 
\unit[0.6]{m} $<~\Delta$S $<$ \unit[4.5]{m} & 1.0$\pm$0.1   & -0.14$\pm$0.04  & 362.2$\pm$3.3 & 27.6$\pm$8.9\\
\unit[4.5]{m} $<~\Delta$S $<$ \unit[8.0]{m} & 0.47$\pm$0.14 & 0.02$\pm$0.04   & 354.6$\pm$9.1 & 78.9$\pm$17.3\\
$\Delta$S $>$ \unit[8.0]{m}                 & 2.0$\pm$0.1   & 0.01$\pm$0.04   & 363.7$\pm$1.8 & 184.8$\pm$6.5\\ \hline
Single Muons                            & 1.27$\pm$0.01 & 0.013$\pm$0.001 & 364.4$\pm$0.3 & 183.0$\pm$0.9\\ \hline
            \multicolumn{5}{c}{MINOS ND } \\ \hline
$\Delta$S $>$ \unit[0.6]{m}               & 2.51$\pm$0.09 & 0.35$\pm$0.03 & 367.4$\pm$1.3 & 23.7$\pm$2.3\\ 
\unit[0.6]{m} $<~\Delta$S $<$ \unit[1.8]{m} & 2.35$\pm$0.17 & 0.25$\pm$0.05 & 369.0$\pm$2.5 & 26.2$\pm$4.2\\
\unit[1.8]{m} $<~\Delta$S $<$ \unit[3.0]{m} & 2.53$\pm$0.17 & 0.41$\pm$0.05 & 369.3$\pm$2.3 & 25.1$\pm$4.0\\
$\Delta$S $>$ \unit[3.0]{m}                 & 2.64$\pm$0.17 & 0.39$\pm$0.05 & 365.8$\pm$2.1 & 22.1$\pm$3.8\\ \hline
Single Muons                            & 0.268$\pm$0.004 & 0.0116$\pm$0.001 & 365.7$\pm$0.4 & 198.6$\pm$0.9\\ \hline \hline
\end{tabular}
\caption{The parameters obtained when Eq.~\ref{eq:fit} 
is fit to the single-muon
and multiple-muon data in each detector.  The table also shows the
results of fits to subsets of the multiple-muon data,
based on the minimum separation between tracks. 
The best fit phase and period do not change significantly 
if the loss rate is assumed to be zero.}
\label{tab:MultiModulation}
\end{table*}

\subsection{Modulations in the Far Detector}

\begin{figure}[thb]
  \begin{center}
    \includegraphics[width=0.48\textwidth]{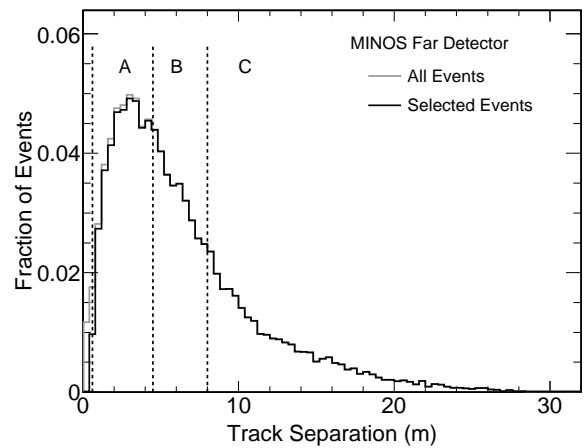}
  \end{center}
  \caption{The minimum track separation $\Delta S$
between any two tracks in multiple-muon 
events recorded in the FD. The gray~(black) histogram is the 
distribution before (after) the selection to remove misreconstructed 
single-muon events. 
Regions of track separation $\Delta S$
are defined as A: \unit[0.6-4.5]{m}, B: \unit[4.5-8.0]{m} and
C: $>$ \unit[8]{m}.}
  \label{fig:farsep}
\end{figure}

The fit for seasonal variations in the FD multiple-muon sample shows a much
smaller amplitude than for single muons, and a poorly defined phase.
Since the 
MINOS FD is larger than the ND and is fully instrumented, the
modulation is studied as a function of 
track separation.
Figure~\ref{fig:farsep} shows the track separation $\Delta S$.
The multiple-muon data are 
grouped into three bins of roughly equal statistics with track 
separations from \unit[0.6-4.5]{m} (FD region A), \unit[4.5-8.0]{m} 
(FD region B) and greater than 
\unit[8]{m} (FD region C). Region A most closely
resembles the distribution in the ND.

\begin{figure}[thb]
  \begin{center}
    \includegraphics[width=0.53\textwidth]{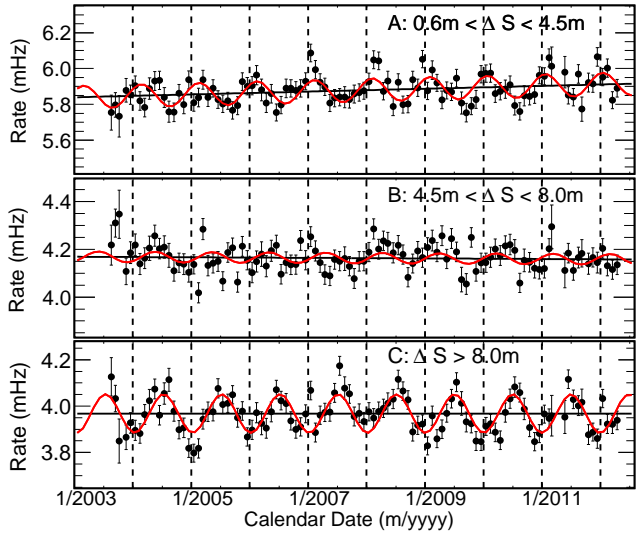}
  \end{center}
  \caption{The multiple-muon rate in the FD as a function of 
time for different track separations.
Each data point 
corresponds to one calendar month of data.  The solid red lines are the 
best fit to Eq.~\ref{eq:fit}.  The top graph is for the smallest 
track separation, the middle graph for mid-range and 
the bottom graph for the largest.  
The vertical lines are year boundaries and
the solid horizontal line represents the fit without the cosine term.}
  \label{fig:FarMultiMuonsVsTime}
\end{figure}

Figure~\ref{fig:FarMultiMuonsVsTime} presents the multiple-muon rate in the 
MINOS FD as a function of time for differing track 
separations.  
The FD multiple-muon data 
set with the largest track separation, \unit[$\>$8]~{m}, modulates 
with a summer maximum~($t_{0}=~$\unit[184.8$\pm$6.5]{days}); this phase 
is consistent with that observed in the FD single-muon sample, and the
amplitude is larger.
On the other hand,
the FD multiple-muon data set with the smallest track 
separations modulates with a winter maximum~($t_{0}=~$\unit[27.6$\pm$8.9]{days}); 
this phase differs by a half year from the variation seen with single muons.
The FD mid-range track 
separation multiple-muon data set 
has a small amplitude and is consistent with an admixture of the
other two phases.

\begin{figure}[thb]
  \begin{center}
    \includegraphics[width=0.48\textwidth]{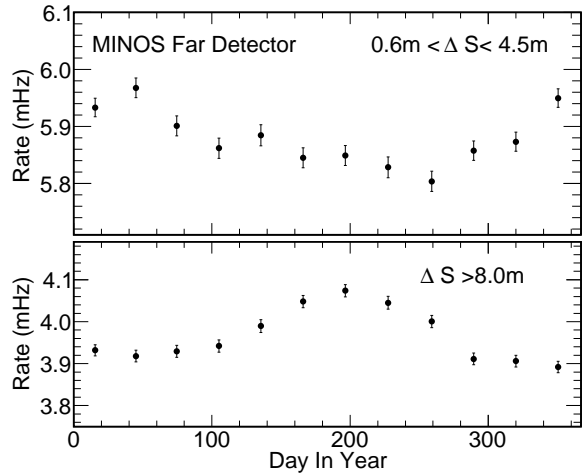}
  \end{center}
  \caption{The multiple-muon rate in the FD for events with
$\Delta S$ range A from 0.6 m to 4.5 m (top graph) 
and for events with $\Delta S$ range C larger than 8 m 
(bottom) binned according to calendar month.
The top figure shows a winter maximum.
The bottom figure shows a summer maximum.}
  \label{fig:farfolded}
\end{figure}

In Fig.~\ref{fig:farfolded}, the data for regions A and C have been
binned by calendar month, with each point showing the average rate over
all years of data-taking.

\subsection{Modulations in the Near Detector}
\begin{figure}[thb]
  \begin{center}
    \includegraphics[width=0.48\textwidth]{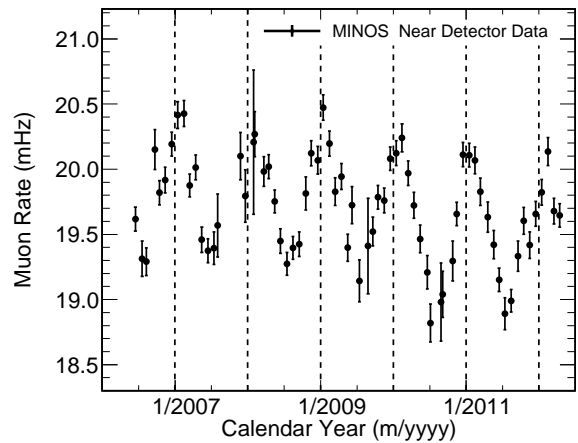}
  \end{center}
  \caption{The multiple-muon rate in the ND as a function of 
time. Each data point corresponds to one calendar month. A clear 
modulation in the data is observed with the maximum occurring towards 
the start of the year. The vertical lines are year boundaries.}
  \label{fig:MultiMuonsVsTime}
\end{figure}

\begin{figure}[thb]
  \begin{center}
    \includegraphics[width=0.53\textwidth]{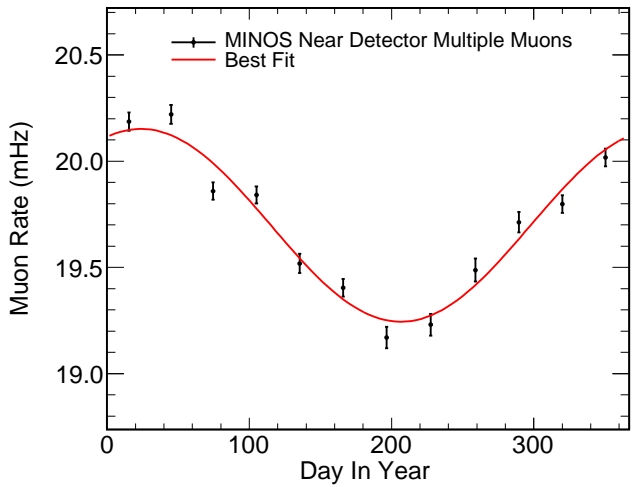}
    \includegraphics[width=0.53\textwidth]{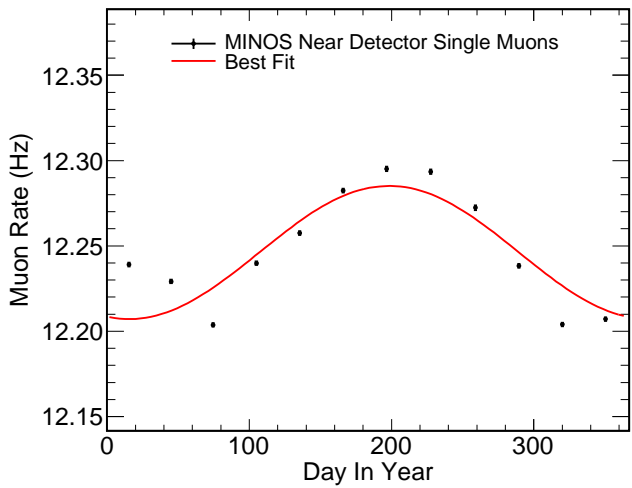}
  \end{center}
  \caption{
The top figure is the multiple-muon rate in the ND, binned according to calendar month,
which each point showing the average rate for all years of data-taking.
The figure also shows a cosine fit to the data. The single-muon
rate is shown in the bottom figure, showing a clearly different seasonal
modulation.}
  \label{fig:nearfolded}
\end{figure}

The ND multiple-muon data, shown in 
Fig.~\ref{fig:MultiMuonsVsTime}, and the single-muon data~(shown in 
Reference~\cite{Adamson:2014})
were fit to Eq.~\ref{eq:fit}
using one month time
interval bins.  
The multiple-muon event 
rate data show a clear modulation signature. However, unlike the 
single muon rate which reaches its maximum in the 
summer~\cite{Adamson:2014}, the multiple-muon rate reaches its maximum 
in the winter.  This also matches the modulation for the region-A
multiple muons in the FD.
Both the single-muon and 
multiple-muon data sets have periods consistent with one year but their 
phases, \unit[198.6$\pm$0.9]{days} and \unit[23.7$\pm$2.3]{days} 
respectively, 
differ by about six months.
The rates of multiple muons and single muons, binned by calendar month
and averaged over all years of data-taking, are
shown in 
Fig.~\ref{fig:nearfolded}.

\begin{figure}[thb]
  \begin{center}
    \includegraphics[width=0.48\textwidth]{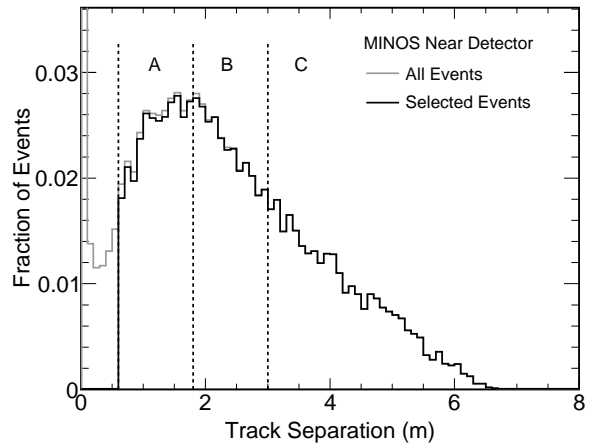}
  \end{center}
  \caption{The minimum track separation $\Delta S$
between any two tracks in multiple-muon
events recorded in the ND. The gray~(black) histogram is the
distribution before(after) the selection to remove misreconstructed
single-muon events. 
Regions of track separation $\Delta S$
are defined as A: \unit[0.6-1.8]{m}, B: \unit[1.8-3.0]{m} and
C: $>$ \unit[3]{m}.}
  \label{fig:nearsep}
\end{figure}

Figure \ref{fig:nearsep} shows the track separation 
in ND multiple-muon events.  To qualitatively match the
procedure in the FD, the data have been grouped 
into three bins of roughly equal statistics with 
track separations of 
\unit[0.6-1.8]{m} (ND region A), \unit[1.8-3.0]{m} (ND region B)
and greater than \unit[3]{m} (ND region C). 
As before, the 
data are fit to Eq.~(\ref{eq:fit}) and the best fit  
parameters are given in Table~\ref{tab:MultiModulation}. 
There is no apparent difference in the fit parameters 
for the three ND regions, which all peak in the winter.   
There is consistency between ND regions ABC and FD region A
in both $\Delta S$ and a winter maximum.

\begin{figure}[thb]
  \begin{center}
    \includegraphics[width=0.48\textwidth]{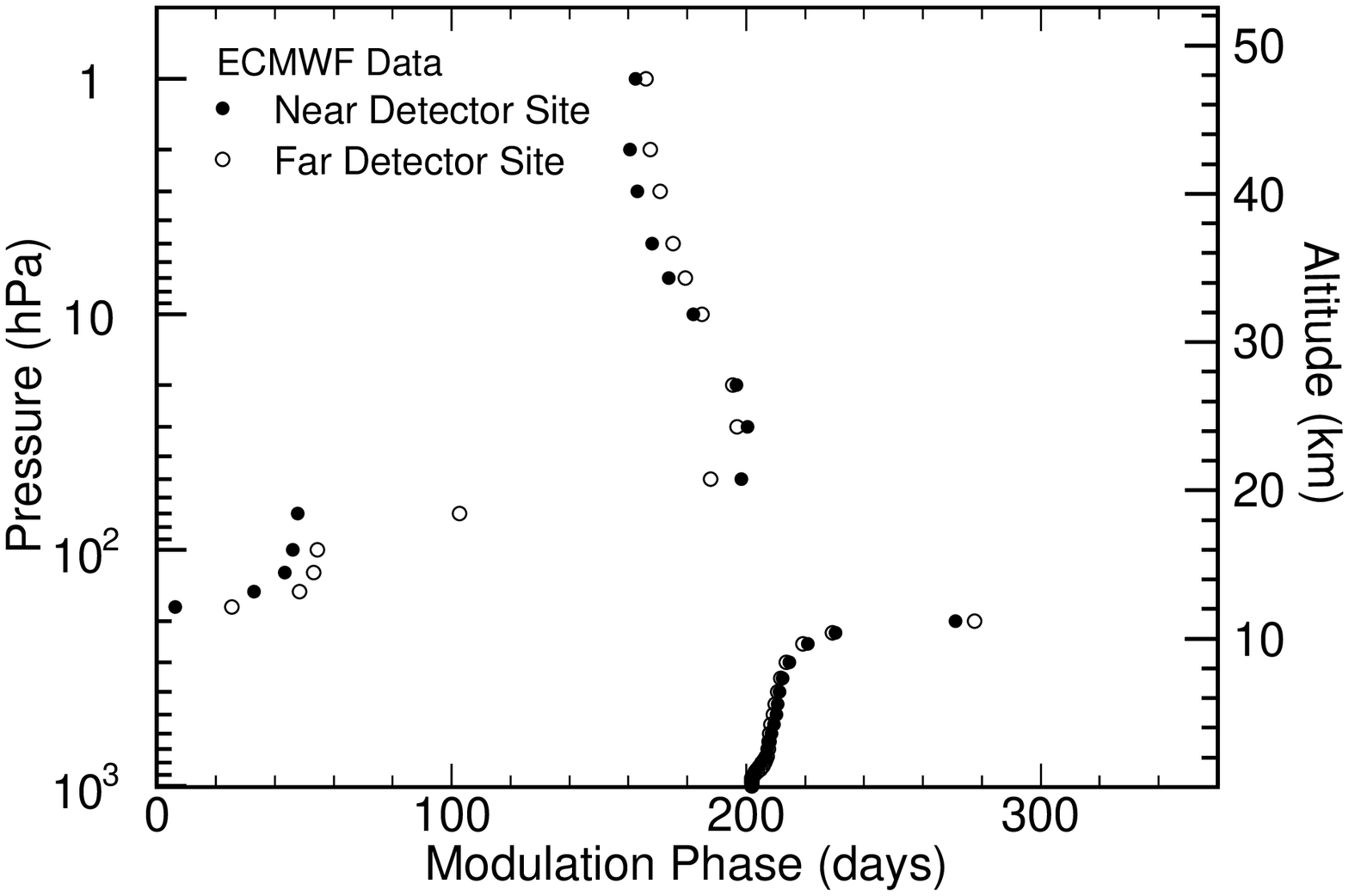}
    \includegraphics[width=0.48\textwidth]{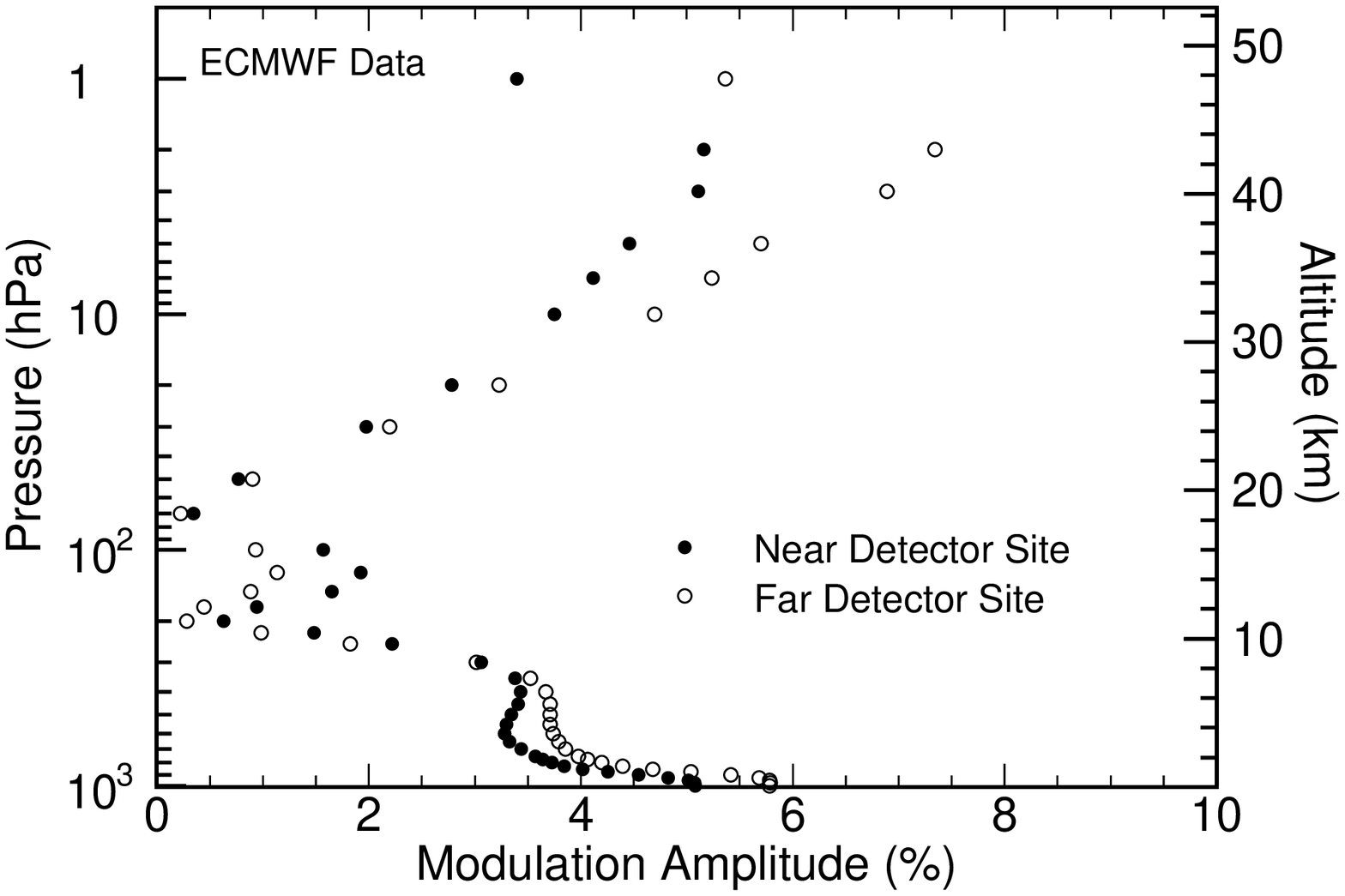}
  \end{center}
  \caption{ The
(top) modulation phase and (bottom) amplitude in the 
ECMWF temperature data based on a cosine fit
are shown as a function of altitude and detector site.
These distributions were used to study both the geometry effect (B) and
the temperature effect (C).  }
  \label{fig:ecmwf}
\end{figure}

\section{Discussion of Results and Possible Explanations}
\label{sec:Discussion}

We have previously observed seasonal variations in single-muon
rates in the MINOS ND and FD that correlate at expected
levels with the temperature changes and the season.  Those muon rates
rose in the summer as did the calculated values of $T_{eff}$, and the
measured correlations were $\at^{ND}$ = 0.428 $\pm$ 0.059 \cite{Adamson:2014}
and $\at^{FD}$ = 0.873 $\pm$ 0.014 \cite{Adamson:2009zf}.  The measurement
of a multiple-muon rate in the ND that peaks in the winter is
unexpected, as is the winter maximum in the 
FD in region A
of separation.  In order to try to understand this result, 
four plausible explanations which might account for these results
are considered.
They involve: 
A) a
source of dimuons from prompt hadron decays 
(such as $\eta$ and $\rho$)
that may have the opposite
seasonal variation, since in the winter the secondary pions are more 
likely to interact than decay and produce more of such hadrons;
B) a geometric effect in which different altitude distributions
affect the track separation underground;
C) a different altitude distribution for multimuon
events that may come from regions of the atmosphere with different
seasonal temperature profiles;
and D) 
leading secondary hadrons being more likely to  decay than interact in
the summer, and thus less likely to make multiple hadrons 
which make multiple muons. 
We discuss each of these possibilities
in the current section.


\subsection{Hadronic dimuon decays}
One idea is that the winter maximum  may be due to hadronic decays
into dimuons.  In the
winter, while pions are less likely to decay in the atmosphere, the
decay probability of other hadrons which have dimuon decays, such
as $\eta$ and $\rho$ mesons, changes negligibly.  The 2\% more pions
\cite{Adamson:2009zf}
which interact will increase the number of these other hadrons.
This increase, which is at most 2\%,
must then be folded in with the small dimuon branching ratios,
such as 4.6 $\times~10^{-5}$ for $\rho \rightarrow \mu^+ \mu^-$ and
3.1 $\times~10^{-4}$ for $\eta \rightarrow \mu^+ \mu^- \gamma$
\cite{pdg:2012}.  Observed dimuon rates are 1\% of the single muon
rates in the FD, and 0.16\% in the ND, so even if $\rho$ and $\eta$
production were comparable to $\pi$, this contribution is at most
6 $\times~10^{-6}$, too small to account for the observed effect.

\subsection{A geometry effect}
\label{sec:geo}
A possibility is that the muons generated higher
in the atmosphere in the summer spread out farther so that there
are fewer of them in region A.  This would be solely a geometric
effect, in that it would not affect the number of multimuons
in each season but only the track-separation distribution.  
This is further complicated by multiple scattering, but an effect
due to the opening angle at production can be estimated.
For a fixed-size
detector, a difference in the track separation distribution would affect
the measured rate.  
The altitude of the first interaction
in an isothermal atmosphere is related to 
the absolute temperature.  A $\pm$ 2\% seasonal change in the effective 
temperature would
cause a $\pm$ 2\% change in the altitude, and hence less than a 4\%
change in the average muon track separation underground.  
This would move events to the right in Fig.~\ref{fig:farsep}.
Due to the shape of the distribution, more events would move
from region A to region B than from region B to region C, which is
in contradiction to our fits.
Also, one would expect a similar effect in the ND as shown in
Fig.~\ref{fig:nearsep}, but no track separation dependence is
seen in the ND.

\subsection{A temperature effect}
\label{sec:temp}
To determine whether there may be an altitude-dependent seasonal
variation that differs for single and multiple muons, meteorological
data is used to determine the atmospheric temperature profile.
Figure~\ref{fig:ecmwf} gives the phase and amplitude of the 
modulation of the atmospheric temperature,
based on a cosine fit to data taken from the European 
Center for Medium-Range Weather Forecasts~(ECMWF) model \cite{ECMWF}, 
as a function of atmospheric pressure. 
Indeed, there is a small region of the 
atmosphere, between \unit[70]{hPa} and \unit[175]{hPa}, where the 
temperature reaches a maximum in the winter. 
Note, however, the small amplitude of the annual temperature variation
at those altitudes.  

In order to study the possible altitude dependence of multiple muons
we simulated cosmic ray air showers which
could make multiple muons in the MINOS FD.
The Monte Carlo sample was produced by CORSIKA 
\cite{Heck:2010, Heck:1998} using version 7.4.
We have run CORSIKA
with three different hadronic 
models, QGSJET-01C, QGSJET II-04 \cite{Ostapchenko:2011}
and EPOS \cite{Werner:2006} which gave consistent results.  
We note that CORSIKA uses
an isothermal atmosphere and cannot be used per se to
study seasonal variations.  The goal here is to roughly calculate the
altitude dependence for the three regions of track separation.
CORSIKA outputs muon energies and positions at the earth's surface.
To reach the MINOS FD,  
energy loss through the rock was calculated using \cite{reichenbacher:2008}:
\begin{equation}
	E_{loss}(X) = \frac{a}{b_T} (e^{b_T X} - 1),
\end{equation}
where $X$ is the rock overburden, $a$ is a parameter for the 
ionization energy loss and $b_T = b_{brem} + b_{pair} + b_{DIS}$ 
represents the energy loss due to bremsstrahlung, electron-positron
pair production and 
photo-nuclear interactions. 

\begin{figure}[thb]
  \begin{center}
    \includegraphics[width=0.48\textwidth]{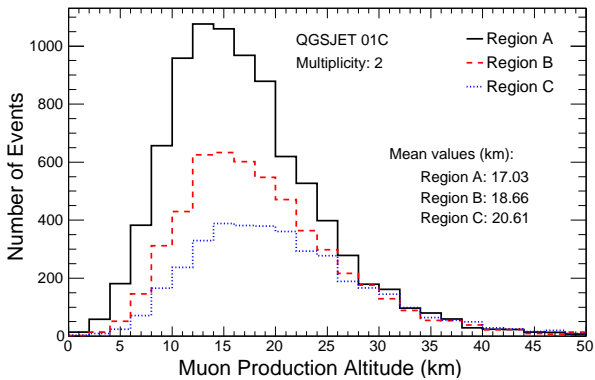}
  \end{center}
  \caption{To study a possible temperature effect with altitude, 
(Sec.~\ref{sec:temp}  in the text),
the altitude distribution from CORSIKA for MINOS FD
multiple muons are shown for each of the three regions of
track separation in Fig.~\ref{fig:farsep}.}
  \label{fig:alt}
\end{figure}

Simulated events were selected for which two or more muons reached the
top of the FD with a total remaining energy of at least 0.9 GeV.
The 
distribution of track separation obtained with this
simulation was similar to, but not 
identical to, the distribution seen in data (Fig.~\ref{fig:farsep}).
We then extracted from CORSIKA the altitude at which each muon was created
in each track-separation region.
Those three distributions are shown in 
Fig.~\ref{fig:alt}.  There is a shift in the mean
altitude for each region of track separation from \mbox{17 km} in region A to
21 km in region C, though all three distributions are quite
broad.
We then combined the altitude dependence with the temperature phase
and amplitude fits shown in Fig.~\ref{fig:ecmwf},
assuming the rate and $T_{eff}$ were completely correlated,
 to compare the
overall variation of $T_{eff}$ averaged over each track-separation
region.  The result was a variation that peaked in the summer
in all three regions, with an amplitude of 1.9\% in region C
and 1.6\% in regions A and B.  This study was repeated using
QGSJET-01C, QGSJET II and EPOS and all three results were similar.
It does not appear that the temperature variations noted in
Fig.~\ref{fig:ecmwf} can account for the observed reverse
seasonal effect in region A.  

\subsection{Anticorrelation of primary and secondary decays}
As a last hypothesis, while most single-muon
events come from secondary pions and kaons produced in the primary cosmic
ray interaction, multiple muons may be more likely to come from
higher energy primaries where there are further hadronic 
interactions deeper in the shower.  In that case, if the secondary
hadron is more likely to decay in the summer, it is less likely to 
interact and make additional
pions and kaons which contribute to multiple
muons.  
This may be the best
explanation for the winter maximum measured in the MINOS ND
multiple-muon data set.  A quantitative test of this hypothesis
will require a detailed study of air shower development that is
beyond the scope of this analysis.
This hypothesis accounts for the stronger
effect in the MINOS ND, where the muons come from
pions and kaons below their critical energies ($\epsilon_\pi$ = 115 GeV
and $\epsilon_K$ = 850 GeV, defined as those energies for which meson
decay and interaction rates in the atmosphere where muons originate
are equal) \cite{Barret:1952} 
and for the more complex effect in the 
MINOS FD where the energies are above $\epsilon_\pi$ and
comparable to $\epsilon_K$.  Mesons which are much below their
critical energies mostly decay, so the temperature effect that
does exist to increase the decay rate in the summer has a large 
effect on decreasing the interaction rate in the summer.  This
is the situation for muons in the ND where the threshold from the
overburden is near 50 GeV.  At the FD, where the threshold is almost
a TeV, a change in the decay rate has a smaller impact on the 
interaction rate, since a large fraction of the hadrons are
interacting before they decay.
\par As pointed out in the introduction, single muons come predominantly
from the decay of a leading hadron, and multiple muons from a more
complicated process.  It is clear that if a leading hadron is more 
likely to decay in one season, it is less likely to interact.

\section{Conclusion}
\label{sec:Conclusion}
We have shown evidence of an annual modulation in the MINOS 
ND multiple-muon data set in which the maximum rate occurs 
in the winter. This phase is inconsistent with 
the summer maximum observed in the ND and FD  single-muon data. Data 
collected by the MINOS FD were used to show that there is a  
transition from a summer maximum in multiple-muon events with
a large track separation to a winter maximum in multiple-muon
events with a small track separation.  This transition
occurs at 
track separations of about  \unit[5-8]{m}. 

Four possible explanations for this observed
characteristic in seasonal 	
variations were considered. 
One explanation is favored: this is a hypothesis in which 
multiple muons come preferentially from higher energy pions and kaons 
which, in the summer, are less likely to interact and produce the 
secondary pions and kaons that give rise to the multiple muons. 	
However, a full explanation of our observations including the
dependence in the FD on track separation must come from a more 
detailed study of extensive air-shower properties and the 
properties of the atmosphere.

\section{Acknowledgments}
\label{sec:Acknowledgements} 
This work was supported by the US DOE, the United Kingdom STFC, 
the US NSF, the state and University of Minnesota, and Brazil's 
FAPESP, CNPq and CAPES. We are grateful to the 
Minnesota Department of Natural Resources and the personnel of 
the Soudan Laboratory and
Fermilab for their contributions to the experiment. 
Fermilab is operated by Fermi Research Alliance, LLC under Contract No.
De-AC02-07CH11359 with the United States Department of Energy.
\bibliography{SeasonalPaperMulti}

\begin{thebibliography}{26}
\expandafter\ifx\csname natexlab\endcsname\relax\def\natexlab#1{#1}\fi
\expandafter\ifx\csname bibnamefont\endcsname\relax
  \def\bibnamefont#1{#1}\fi
\expandafter\ifx\csname bibfnamefont\endcsname\relax
  \def\bibfnamefont#1{#1}\fi
\expandafter\ifx\csname citenamefont\endcsname\relax
  \def\citenamefont#1{#1}\fi
\expandafter\ifx\csname url\endcsname\relax
  \def\url#1{\texttt{#1}}\fi
\expandafter\ifx\csname urlprefix\endcsname\relax\def\urlprefix{URL }\fi
\providecommand{\bibinfo}[2]{#2}
\providecommand{\eprint}[2][]{\url{#2}}

\bibitem[{\citenamefont{Barret et~al.}(1952)}]{Barret:1952}
\bibinfo{author}{\bibfnamefont{P.}~\bibnamefont{Barret}} \bibnamefont{et~al.},
  \bibinfo{journal}{Rev. Mod. Phys.} \textbf{\bibinfo{volume}{24}},
  \bibinfo{pages}{133} (\bibinfo{year}{1952}).

\bibitem[{\citenamefont{Castagnoli and Dodero}(1967)}]{Castagnoli:1967}
\bibinfo{author}{\bibfnamefont{G.}~\bibnamefont{Castagnoli}} \bibnamefont{
and M. A. Dodero},
  \bibinfo{journal}{Il Nuovo Cim. B} \textbf{\bibinfo{volume}{51}},
  \bibinfo{pages}{525} (\bibinfo{year}{1967}).

\bibitem[{\citenamefont{Sherman}(1954)}]{Sherman:1954}
\bibinfo{author}{\bibfnamefont{N.}~\bibnamefont{Sherman}},
  \bibinfo{journal}{Phys. Rev.} \textbf{\bibinfo{volume}{93}},
  \bibinfo{pages}{208} (\bibinfo{year}{1954}).

\bibitem[{\citenamefont{Castagnoli and Dodero}(1952)}]{Torino:1967}
\bibinfo{author}{\bibfnamefont{G.~C.} \bibnamefont{Castagnoli}}
  \bibnamefont{and} \bibinfo{author}{\bibfnamefont{M.}~\bibnamefont{Dodero}}
  (\bibinfo{collaboration}{Torino Collaboration}), 
\bibinfo{journal}{Rev. Mod. Phys.}
  \textbf{\bibinfo{volume}{24}}, \bibinfo{pages}{133} (\bibinfo{year}{1952}).

\bibitem[{\citenamefont{Fenton et~al.}(1961)\citenamefont{Fenton, Jacklyn, and
  Taylor}}]{Hobart:1961}
\bibinfo{author}{\bibfnamefont{A.}~\bibnamefont{Fenton}},
  \bibinfo{author}{\bibfnamefont{R.}~\bibnamefont{Jacklyn}}, \bibnamefont{and}
  \bibinfo{author}{\bibfnamefont{R.}~\bibnamefont{Taylor}},
  \bibinfo{journal}{Il Nuovo Cim. B} \textbf{\bibinfo{volume}{22}},
  \bibinfo{pages}{285} (\bibinfo{year}{1961}).

\bibitem[{\citenamefont{Andreyev et~al.}(1987)}]{Baksan:1987}
\bibinfo{author}{\bibfnamefont{Y.}~\bibnamefont{Andreyev}} \bibnamefont{et~al.}
  (\bibinfo{collaboration}{Baksan Collaboration}), 
\bibinfo{howpublished}{{Proceedings of the
  20th ICRC}, vol. 3, pp. 270} (\bibinfo{year}{1987}).

\bibitem[{\citenamefont{Ambrosio et~al.}(1997{\natexlab{a}})}]{Ambrosio:1997tc}
\bibinfo{author}{\bibfnamefont{M.}~\bibnamefont{Ambrosio}} \bibnamefont{et~al.}
  (\bibinfo{collaboration}{MACRO Collaboration}), \bibinfo{journal}{Astropart.
  Phys.} \textbf{\bibinfo{volume}{7}}, \bibinfo{pages}{109}
  (\bibinfo{year}{1997}{\natexlab{a}}).

\bibitem[{\citenamefont{Bouchta}(1999)}]{Bouchta:1999}
\bibinfo{author}{\bibfnamefont{A.}~\bibnamefont{Bouchta}}
  (\bibinfo{collaboration}{AMANDA Collaboration}),
  \bibinfo{howpublished}{{Proceedings of the 26th ICRC}, vol. 2, pp. 108-111}
  (\bibinfo{year}{1999}).

\bibitem[{\citenamefont{Bellini et~al.}(2012)}]{Bellini:2012te}
\bibinfo{author}{\bibfnamefont{G.}~\bibnamefont{Bellini}} \bibnamefont{et~al.}
  (\bibinfo{collaboration}{Borexino Collaboration}), \bibinfo{journal}{J. Cosm.
  Astropart. Phys.} \textbf{\bibinfo{volume}{1205}}, \bibinfo{pages}{015}
  (\bibinfo{year}{2012}), \eprint{hep-ex/1202.6403}.

\bibitem[{\citenamefont{Selvi}(2009)}]{Solvi:2009}
\bibinfo{author}{\bibfnamefont{M.}~\bibnamefont{Selvi}}
  (\bibinfo{collaboration}{LVD Collaboration}),
  \bibinfo{howpublished}{{Proceedings of the 31st ICRC}}
  (\bibinfo{year}{2009}).

\bibitem[{\citenamefont{Desiati et~al.}(2011)}]{Desiati:2011}
\bibinfo{author}{\bibfnamefont{P.}~\bibnamefont{Desiati}} \bibnamefont{et~al.}
  (\bibinfo{collaboration}{IceCube Collaboration}),
  \bibinfo{howpublished}{{Proceedings of the 32nd ICRC}}
  (\bibinfo{year}{2011}), \eprint{astro-ph/1111.2735}.

\bibitem[{\citenamefont{Osprey et~al.}(2009)}]{Osprey:2009ni}
\bibinfo{author}{\bibfnamefont{S.}~\bibnamefont{Osprey}} \bibnamefont{et~al.}
  (\bibinfo{collaboration}{MINOS Collaboration}), \bibinfo{journal}{Geophys.
  Res. Lett.} \textbf{\bibinfo{volume}{36}}, \bibinfo{pages}{L05809}
  (\bibinfo{year}{2009}).

\bibitem[{\citenamefont{Adamson et~al.}(2010)}]{Adamson:2009zf}
\bibinfo{author}{\bibfnamefont{P.}~\bibnamefont{Adamson}} \bibnamefont{et~al.}
  (\bibinfo{collaboration}{MINOS Collaboration}), 
\bibinfo{journal}{Phys. Rev.  D} 
  \textbf{\bibinfo{volume}{81}}, \bibinfo{pages}{012001}
  (\bibinfo{year}{2010}), \eprint{hep-ex/0909.4012}.

\bibitem[{\citenamefont{Adamson et~al.}(2014)}]{Adamson:2014}
\bibinfo{author}{\bibfnamefont{P.}~\bibnamefont{Adamson}} \bibnamefont{et~al.}
  (\bibinfo{collaboration}{MINOS Collaboration}) 
\bibinfo{journal}{Phys. Rev.  D} 
  \textbf{\bibinfo{volume}{90}}, \bibinfo{pages}{012010}
  (\bibinfo{year}{2014}).

\bibitem[{\citenamefont{Michael et~al.}(2008)}]{Michael:2008bc}
\bibinfo{author}{\bibfnamefont{D.~G.} \bibnamefont{Michael}}
  \bibnamefont{et~al.} (\bibinfo{collaboration}{MINOS Collaboration}),
  \bibinfo{journal}{Nucl. Instrum. Meth.} \textbf{\bibinfo{volume}{A596}},
  \bibinfo{pages}{190} (\bibinfo{year}{2008}),
  \eprint{physics.ins-det/0805.3170}.

\bibitem[{\citenamefont{Adamson et~al.}}()]{bib:cesar}
\bibinfo{author}{\bibfnamefont{P.}~\bibnamefont{Adamson}} 
  \bibnamefont{et~al.} (\bibinfo{collaboration}{MINOS Collaboration}),
\bibinfo{note}{The Multiple Muon Charge Ratio in MINOS, in preparation}.

\bibitem[{\citenamefont{Kasahara et~al.}(2008)}]{bib:kasahara}
\bibinfo{author}{\bibfnamefont{S.} \bibnamefont{Kasahara}}
  \bibnamefont{et~al.} (\bibinfo{collaboration}{Soudan 2}),
  \bibinfo{journal}{Phys.Rev.} \textbf{\bibinfo{volume}{D55}},
  \bibinfo{pages}{5282} (\bibinfo{year}{1997}),

\bibitem[{\citenamefont{Maciuc et~al.}(2006)\citenamefont{Maciuc, Grupen,
  Hashim, Luitz, Mailov et~al.}}]{Maciuc:2006xb}
\bibinfo{author}{\bibfnamefont{F.}~\bibnamefont{Maciuc}}
  \bibnamefont{et~al.}, \bibinfo{journal}{Phys. Rev. Lett.}
  \textbf{\bibinfo{volume}{96}}, \bibinfo{pages}{021801}
  (\bibinfo{year}{2006}).

\bibitem[{\citenamefont{Kudryavtsev et~al.}(1999)\citenamefont{Kudryavtsev,
  Korolkova, and Spooner}}]{Kudryavtsev:1999zu}
\bibinfo{author}{\bibfnamefont{V.}~\bibnamefont{Kudryavtsev}},
  \bibinfo{author}{\bibfnamefont{E.}~\bibnamefont{Korolkova}},
  \bibnamefont{and} \bibinfo{author}{\bibfnamefont{N.}~\bibnamefont{Spooner}},
  \bibinfo{journal}{Phys. Lett.} \textbf{\bibinfo{volume}{B471}},
  \bibinfo{pages}{251} (\bibinfo{year}{1999}), \eprint{hep-ph/9911493}.

\bibitem[{\citenamefont{Ambrosio et~al.}(1999)}]{Ambrosio:1999qu}
\bibinfo{author}{\bibfnamefont{M.}~\bibnamefont{Ambrosio}} \bibnamefont{et~al.}
  (\bibinfo{collaboration}{MACRO Collaboration}), \bibinfo{journal}{Phys. Rev.}
  \textbf{\bibinfo{volume}{D60}}, \bibinfo{pages}{032001}
  (\bibinfo{year}{1999}), \eprint{hep-ex/9901027}.

\bibitem[{\citenamefont{Berger et~al.}(1989)}]{Berger:1989hs}
\bibinfo{author}{\bibfnamefont{C.}~\bibnamefont{Berger}} \bibnamefont{et~al.}
  (\bibinfo{collaboration}{Frejus Collaboration}), \bibinfo{journal}{Phys. Rev.}
  \textbf{\bibinfo{volume}{D40}}, \bibinfo{pages}{2163} (\bibinfo{year}{1989}).

\bibitem[{\citenamefont{Anzivino et~al.}(1990)\citenamefont{Anzivino, Bianco,
  Casaccia, Cindolo, De~Felice et~al.}}]{Anzivino:1990xi}
\bibinfo{author}{\bibfnamefont{G.}~\bibnamefont{Anzivino}}
  \bibnamefont{et~al.}, \bibinfo{journal}{Nucl. Instrum. Meth.}
  \textbf{\bibinfo{volume}{A295}}, \bibinfo{pages}{466} (\bibinfo{year}{1990}).

\bibitem[{\citenamefont{Ambrosio et~al.}(1997{\natexlab{b}})}]{Ambrosio:1996et}
\bibinfo{author}{\bibfnamefont{M.}~\bibnamefont{Ambrosio}} \bibnamefont{et~al.}
  (\bibinfo{collaboration}{MACRO Collaboration}), \bibinfo{journal}{Phys. Rev.}
  \textbf{\bibinfo{volume}{D56}}, \bibinfo{pages}{1407}
  (\bibinfo{year}{1997}{\natexlab{b}}).

\bibitem[{\citenamefont{Kasahara et~al.}(1997)}]{Kasahara:1996kw}
\bibinfo{author}{\bibfnamefont{S.}~\bibnamefont{Kasahara}} \bibnamefont{et~al.}
  (\bibinfo{collaboration}{Soudan 2 Collaboration}),
  \bibinfo{journal}{Phys. Rev.} \textbf{\bibinfo{volume}{D55}},
  \bibinfo{pages}{5282} (\bibinfo{year}{1997}), \eprint{hep-ex/9612004}.

\bibitem[{\citenamefont{Avati et~al.}(2003)\citenamefont{Avati, Dick, Eggert,
  Strom, Wachsmuth et~al.}}]{Avati:2000mn}
\bibinfo{author}{\bibfnamefont{V.}~\bibnamefont{Avati}}
  \bibnamefont{et~al.}, \bibinfo{journal}{Astropart. Phys.}
  \textbf{\bibinfo{volume}{19}}, \bibinfo{pages}{513} (\bibinfo{year}{2003}).

\bibitem[{\citenamefont{Abdallah et~al.}(2007)}]{Abdallah:2007fk}
\bibinfo{author}{\bibfnamefont{J.}~\bibnamefont{Abdallah}} \bibnamefont{et~al.}
  (\bibinfo{collaboration}{DELPHI Collaboration}),
  \bibinfo{journal}{Astropart. Phys.} \textbf{\bibinfo{volume}{28}},
  \bibinfo{pages}{273} (\bibinfo{year}{2007}), \eprint{0706.2561}.

\bibitem[{\citenamefont{Aglietta et~al.}(2004)}]{Aglietta:2003hq}
\bibinfo{author}{\bibfnamefont{M.}~\bibnamefont{Aglietta}} \bibnamefont{et~al.}
  (\bibinfo{collaboration}{MACRO Collaboration, EAS-TOP Collaboration}),
  \bibinfo{journal}{Astropart. Phys.} \textbf{\bibinfo{volume}{20}},
  \bibinfo{pages}{641} (\bibinfo{year}{2004}), \eprint{astro-ph/0305325}.

\bibitem[{\citenamefont{Grashorn}(2008)}]{Grashorn:2008dis}
\bibinfo{author}{\bibfnamefont{E.~W.} \bibnamefont{Grashorn}}, Ph.D. thesis,
  \bibinfo{school}{University of Minnesota} (\bibinfo{year}{2008}),
  \bibinfo{note}{{F}ERMILAB-THESIS-2008-06}.

\bibitem[{ECM(available from http://badc.nerc.ac.uk/data/ecmwf-op/)}]{ECMWF}
\bibinfo{journal}{European Centre for Medium-Range Weather Forecasts ECMWF
  Operational Analysis data, [Internet] British Atmospheric Data Centre}
  \textbf{\bibinfo{volume}{2006-2007}} (\bibinfo{year}{Available from
  http://badc.nerc.ac.uk/data/ecmwf-op/}).


\bibitem[{\citenamefont{Heck}(2010)}]{Heck:2010}
\bibinfo{author}{\bibfnamefont{D.}~\bibnamefont{Heck}} \bibnamefont{et~al.}
(\bibinfo{year}{2010}),
  \bibinfo{note}{Extensive Air Shower Simulation with CORSIKA: 
A User's Guide. Avaliable at: 
www-ik.fzk.de/corsika/usersguide/usersguide.pdf}.




%

\bibitem[{\citenamefont{Heck}(1998)}]{Heck:1998}
\bibinfo{author}{\bibfnamefont{D.}~\bibnamefont{Heck}} \bibnamefont{et~al.}
(\bibinfo{year}{1998}),
  \bibinfo{note}{CORSIKA: A Monte Carlo Code to Simulate 
Extensive Air Showers. Report FZKA 6019}.

\bibitem[{\citenamefont{Ostapchenko}(2011)}]{Ostapchenko:2011}
\bibinfo{author}{\bibfnamefont{S.}~\bibnamefont{Ostapchenko}}, 
  \bibinfo{journal}{Phys. Rev. D} \textbf{\bibinfo{volume}{83}},
  \bibinfo{pages}{14018} (\bibinfo{year}{2011}).

\bibitem[{\citenamefont{Werner}(2006)}]{Werner:2006}
\bibinfo{author}{\bibfnamefont{K.}~\bibnamefont{Werner}} \bibnamefont{et~al.}
  \bibinfo{journal}{Phys. Rev. C} \textbf{\bibinfo{volume}{74}},
  \bibinfo{pages}{44902} (\bibinfo{year}{2006}).

\bibitem[{\citenamefont{Reichenbacher}(2008)}]{reichenbacher:2008}
\bibinfo{author}{\bibfnamefont{J.}~\bibnamefont{Reichenbacher}} 
\bibinfo{author}{\bibfnamefont{and M.}~\bibnamefont{Goodman},} 
Differences in dE/dX for $\mu^+$ and $\mu^-$ and its Effect 
on the Underground Charge Ratio,
\bibinfo{howpublished}{{Proceedings of the
  30th ICRC}} (\bibinfo{year}{2008}).

\bibitem[{\citenamefont{Beringer et~al.}(2012)}]{pdg:2012}
\bibinfo{author}{\bibfnamefont{J.}~\bibnamefont{Beringer}} 
\bibnamefont{et~al.},
  \bibinfo{journal}{Phys. Rev. D} \textbf{\bibinfo{volume}{86}},
  \bibinfo{pages}{010001} (\bibinfo{year}{2012}).

\bibitem[{\citenamefont{S. Tognini}(2012)}]{tognini}
\bibinfo{author}{\bibfnamefont{S.}~\bibnamefont{Tognini}}, 
   M.S. thesis,
  \bibinfo{school}{University of Goias} (\bibinfo{year}{2012}).



\end{thebibliography}

\end{document}